\documentclass[aps,prd,preprint,superscriptaddress,tightenlines,nofootinbib]{revtex4}

\usepackage{graphicx}
\usepackage{dcolumn}
\usepackage{bm}

\usepackage{rotating} 
\usepackage{epsfig}

\newcommand{\rhoe}{\rho^-e^+\nu_e} 
 
\newcommand{\kste}{K^{*-}e^+\nu_e} 
\newcommand{\ke}{K^-e^+\nu_e} 
\newcommand{\pie}{\pi^-e^+\nu_e} 
\newcommand{\drhoe}{D^0\to \rho^-e^+\nu_e} 
\newcommand{\dkste}{D^0\to K^{*-}e^+\nu_e} 
\newcommand{\dke}{D^0\to K^-e^+\nu_e} 
\newcommand{\dkmu}{D^0\to K^-\mu^+\nu_\mu} 
\newcommand{\dkstce}{D^0\to K^{*-}(K^-\pi^0)e^+\nu_e} 
\newcommand{\dkstne}{D^0\to K^{*-}(K_S^0\pi^-)e^+\nu_e} 
\newcommand{\dpie}{D^0\to \pi^-e^+\nu_e} 
\newcommand{\dpimu}{D^0\to \pi^-\mu^+\nu_\mu}

\newcommand{\dbarkpi}{\overline {D}^{0}\to K^+\pi^-} 
\newcommand{\dbarkpp}{\overline {D}^{0}\to K^+\pi^-\pi^0} 
\newcommand{\dbarkppp}{\overline {D}^{0}\to K^+\pi^-\pi^0\pi^0} 
\newcommand{\dbarkpppc}{\overline {D}^{0}\to K^+\pi^-\pi^+\pi^-} 
\newcommand{\dbarksp}{\overline {D}^{0}\to K^0_S\pi^0}
\newcommand{\dbarkspp}{\overline {D}^{0}\to K^0_S\pi^+\pi^-}
\newcommand{\dbarksppp}{\overline {D}^{0}\to K^0_S\pi^+\pi^-\pi^0}
\newcommand{\dbarkk}{\overline {D}^{0}\to K^-K^+}

\newcommand{\kstce}{K^{*-}(K^-\pi^0)e^+\nu_e} 
\newcommand{\kstne}{K^{*-}(K_S^0\pi^-)e^+\nu_e}

\begin{document}

\preprint{CLNS 05-1906}       
\preprint{CLEO 05-01}         

\title{Absolute Branching Fraction Measurements of  Exclusive 
       $D^{0}$ Semileptonic Decays}

\author{T.~E.~Coan}
\author{Y.~S.~Gao}
\author{F.~Liu}
\affiliation{Southern Methodist University, Dallas, Texas 75275}
\author{M.~Artuso}
\author{C.~Boulahouache}
\author{S.~Blusk}
\author{J.~Butt}
\author{E.~Dambasuren}
\author{O.~Dorjkhaidav}
\author{J.~Li}
\author{N.~Menaa}
\author{R.~Mountain}
\author{R.~Nandakumar}
\author{R.~Redjimi}
\author{R.~Sia}
\author{T.~Skwarnicki}
\author{S.~Stone}
\author{J.~C.~Wang}
\author{K.~Zhang}
\affiliation{Syracuse University, Syracuse, New York 13244}
% ...
\collaboration{CLEO Collaboration} 
\noaffiliation

\date{\today}

\begin{abstract} 
With the first data sample collected by the CLEO-c detector at the 
$\psi(3770)$ resonance we have studied four exclusive semileptonic 
decays of the $D^0$ meson.  Our results include the first observation 
and absolute branching fraction measurement for 
$D^0 \rightarrow \rho^{-} e^+ \nu_e$ and improved measurements of the 
absolute branching fractions for $D^0$ decays 
to $K^- e^+ \nu_e$, $\pi^- e^+ \nu_e$, and $K^{*-} e^+ \nu_e$.
\end{abstract}

\maketitle

The weak-current couplings of quarks within the Standard Model are 
described by the elements of the 
Cabibbo-Kobayashi-Maskawa (CKM) matrix~\cite{ckm}, 
which must be determined experimentally.  Because of
their simplicity, semileptonic decays of hadrons 
provide powerful tools for probing the CKM matrix.  
Interpreting experimental measurements of semileptonic decay 
rates requires precision 
knowledge of form factors that are not easily calculated in quantum 
chromodynamics because of non-perturbative effects.  Form factor 
uncertainties are presently the main limitation in extracting 
$|V_{ub}|$ and $|V_{cb}|$ from semileptonic $B$ 
decays~\cite{Artuso:2004gn}.

Heavy Quark Effective Theory (HQET) relates form factors in charm decays 
to those in bottom decays, and lattice gauge techniques 
calculate form factors in both charm and bottom decays.  
Precision measurements of semileptonic charm decay rates  and form
factors are a principal 
goal of the CLEO-c program at the Cornell Electron Storage Ring~\cite{cleoc}.  
 In this Letter, we report first results on semileptonic 
$D^{0}$ decays from CLEO-c: improved measurements of the branching fractions 
for $D^{0} \to \ke$, $\pie$ and $\kste$, and the first observation 
and branching fraction measurement for the decay $\drhoe$. 
(Charge-conjugate modes are implied throughout this Letter.)  
The data sample used for these measurements consists of an integrated 
luminosity of 55.8~pb$^{-1}$ at the $\psi(3770)$ 
resonance, and  includes about 0.20 million $D^0\overline D^0$ 
events~\cite{cleoc-Dtagging}. The 
same data and analysis technique are used for the branching fraction 
measurements of $D^+$ semileptonic decays in~\cite{Dplus}.

The technique for this analysis, which was first applied by the 
Mark III collaboration~\cite{MkIII} at SPEAR, relies on the purity 
and kinematics of $D {\overline D}$ events produced at the 
$\psi(3770)$. We select events by reconstructing a $\overline {D}^{0}$ 
meson in one of eight hadronic final states: $K^+\pi^-$, 
$K^+\pi^-\pi^0$, $K^+\pi^-\pi^0\pi^0$, 
$K^+\pi^-\pi^+\pi^-$, $K^0_S\pi^0$, $K^0_S\pi^+\pi^-$, 
$K^0_S\pi^+\pi^-\pi^0$, and $K^-K^+$.  Within these tagged events, 
$D^{0}$ semileptonic decays are reconstructed in
the exclusive final states: $\ke$, $\pie$, $\kste$, and $\rhoe$,
where $K^{*-}\to K^{-}\pi^{0}$ or $K^{0}_{S}\pi^{-}$, and
$\rho^{-}\to \pi^{-}\pi^{0}$.  
Separation between signal and background from misidentified or 
missing particles is achieved with the 
kinematic variable $U \equiv E_{\rm miss} - c|\vec{p}_{\rm miss}|$, where 
$E_{\rm miss}$ and $\vec{p}_{\rm miss}$ are the missing energy 
and momentum of the $D$ meson decaying semileptonically.
The efficiency-corrected ratio of tagged events with semileptonic 
decays to the total number of tags gives the absolute branching 
fraction for the exclusive semileptonic decay mode.  
This branching fraction is independent of the luminosity of the 
data and benefits from the cancellation of many systematic 
uncertainties.

The efficient reconstruction of tag events and the clean selection 
of semileptonic decays relies on the power of the CLEO-c 
detector, most components of which were developed for and
used in $B$ meson studies in the CLEO~II and CLEO~III 
experiments~\cite{cleo-c}. 
The tracking system covers a solid angle of 
93\% of $4\pi$ with a six-layer low-mass stereo wire drift 
chamber surrounded by a 47-layer cylindrical (main) drift chamber.  
The main drift chamber provides specific-ionization ($dE/dx$) 
measurements that discriminate between charged pions and kaons.  
Additional hadron identification is provided by a Ring-Imaging 
Cherenkov (RICH) detector covering about 80\% of  4$\pi$.
Identification of positrons and detection 
of neutral pions rely on an electromagnetic calorimeter consisting 
of 7800 cesium iodide crystals and covering 95\% of 4$\pi$.  

\begin{table}
\caption{The tag yields for the eight $\overline {D}^{0}$ decay modes with
statistical uncertainties.} 
\begin{center}
\begin{tabular}{lc}\hline \hline
Tag Mode        &  Tag Yield              \\ \hline 
$\dbarkpi$      & 10223$\pm$109  \\ 
$\dbarkpp$      & 18574$\pm$173  \\
$\dbarkppp$     &  4813$\pm$229  \\
$\dbarkpppc$    & 14767$\pm$145  \\  
$\dbarkspp$     &  4879$\pm$99   \\ 
$\dbarksppp$    &  4299$\pm$195  \\ 
$\dbarksp$      &  1585$\pm$49   \\ 
$\dbarkk$       &   901$\pm$32   \\ \hline 
All Tags        & 60041$\pm$408  \\ \hline \hline 
\end{tabular} 
\end{center}
\label{table}
\end{table}

Details of the criteria for selecting tracks, $\pi^{0}$ and $K^0_S$ 
candidates, and hadronic tags are provided in Ref.~\cite{cleoc-Dtagging}.
The tag selection is based on two variables: 
$\Delta E\equiv E_{D}-E_{\rm beam}$,  the difference between 
the energy ($E_{D}$) of the fully reconstructed $\overline {D}^{0}$ 
candidate and the beam energy ($E_{\rm beam}$),
and  
$M_{\rm bc}\equiv \sqrt{E_{\rm beam}^2/c^4 - | \vec{p}_{D} |^2/c^2}$, 
the beam-constrained mass of the $\overline {D}^{0}$ candidate, 
where $\vec{p}_{D}$ is the measured momentum of the 
$\overline {D}^{0}$ candidate. In case of multiple candidates, 
$\Delta E$ is used to select one $\overline {D}^{0}$ 
candidate for each tag mode, 
and we fit the beam-constrained mass distributions to obtain the tag yields. 
The signal component in these fits consists of a Gaussian and a 
bifurcated Gaussian to account for radiative and other effects. 
The background component is represented by an ARGUS function~\cite{argus}.
The yields of tags in all decay modes are given in 
Table~\ref{table}.  The total number of tags in our data sample is 
approximately 60,000.

The requirement of a fully reconstructed $\overline {D}^{0}$ meson tag greatly 
suppresses background.  After a tag is 
identified, we search for a positron and a set of hadrons 
recoiling against the tag. (Only positrons are used because the CLEO-c 
muon identification system
has poor acceptance in the momentum range characteristic of semileptonic
$D$ decays at the $\psi(3770)$.) 
Positron candidates are selected based 
on a likelihood ratio constructed from three inputs: the ratio of the 
energy deposited in the calorimeter to the measured momentum ($E/p$), 
$dE/dx$, and RICH information.  Positron candidates are required to have
momentum of at least 200~MeV/$c$ and to satisfy the fiducial 
requirement $| \cos{\theta} |$ $<$ 0.90, where $\theta$ is the angle 
between the positron direction and the beam axis.  
The minimum momentum is chosen 
because of backgrounds from low-momentum pions.  
More than 80\% of the positrons from $D^{0}$ semileptonic decays at the 
$\psi(3770)$ 
resonance satisfy these requirements. The efficiency for positron 
identification has been measured  
primarily from radiative Bhabha events.  For the criteria used 
in this analysis, it rises from $\sim 50\%$ at 200~MeV/$c$ to 95\% just above
300~MeV/$c$ and then is roughly constant.  The rates for 
misidentifying charged pions and kaons as positrons have been determined 
with exclusive hadronic decays of $K^0_S$ and $D$ mesons in CLEO-c data. 
Averaged over the full momentum range, the pion and  
kaon misidentification rates are approximately 0.1\%. 
Bremsstrahlung photons are recovered by adding showers
that are within 5$^{0}$ of the positron and are not matched to other particles.

To select the hadronic daughters of a semileptonic $D^0$ decay, charged 
pions and kaons with momenta greater than 50~MeV/$c$ are 
identified with criteria based on $dE/dx$ and RICH information. 
Charged pion and kaon candidates must have $dE/dx$ measurements 
within three standard deviations  (3$\sigma$) of the expected values.
For 
tracks with momenta greater than 700~MeV/$c$, RICH information, if available, 
is combined with $dE/dx$.  The efficiencies ($\sim 95$\% or higher) and 
misidentification rates (no more than a few per cent per track) are 
determined with charged pion and kaon samples from fully reconstructed 
decays of $D^0$ and $D^+$ in CLEO-c data.  

We form $\pi^0$ candidates 
with pairs of photons, each with an energy of at least 30 MeV and 
a shower shape consistent with that expected for a photon.  
The invariant mass of the photon pair must be within $3 \sigma$ 
($\sigma \sim$ 6~MeV/$c^2$) of the known $\pi^0$ mass.
After selection, a mass constraint is imposed when $\pi^0$ candidates are
used in reconstructing other states.
We reconstruct pairs of oppositely charged tracks from a common vertex 
to form a $K_S^0$ candidate within 12 MeV/$c^2$ ($\sim4.5\sigma$) 
of the $K_S^0$ mass.  

The $K^{-}$ and $\pi^{0}$, or $K^0_S$ and $\pi^-$, candidates are combined 
to form $K^{*-}$ candidates. 
We require the 
invariant masses of the $K^{*-}$ candidates to be within 100~MeV/$c^2$ 
of the mean $K^{*-}$ mass. Likewise,
$\pi^{-}$ and $\pi^{0}$ candidates are combined to form 
$\rho^{-}$ candidates within 150~MeV/$c^2$ of the mean $\rho^{-}$ mass. 

\begin{figure}[htbp]
{\mbox{\includegraphics[width=0.50\textwidth]{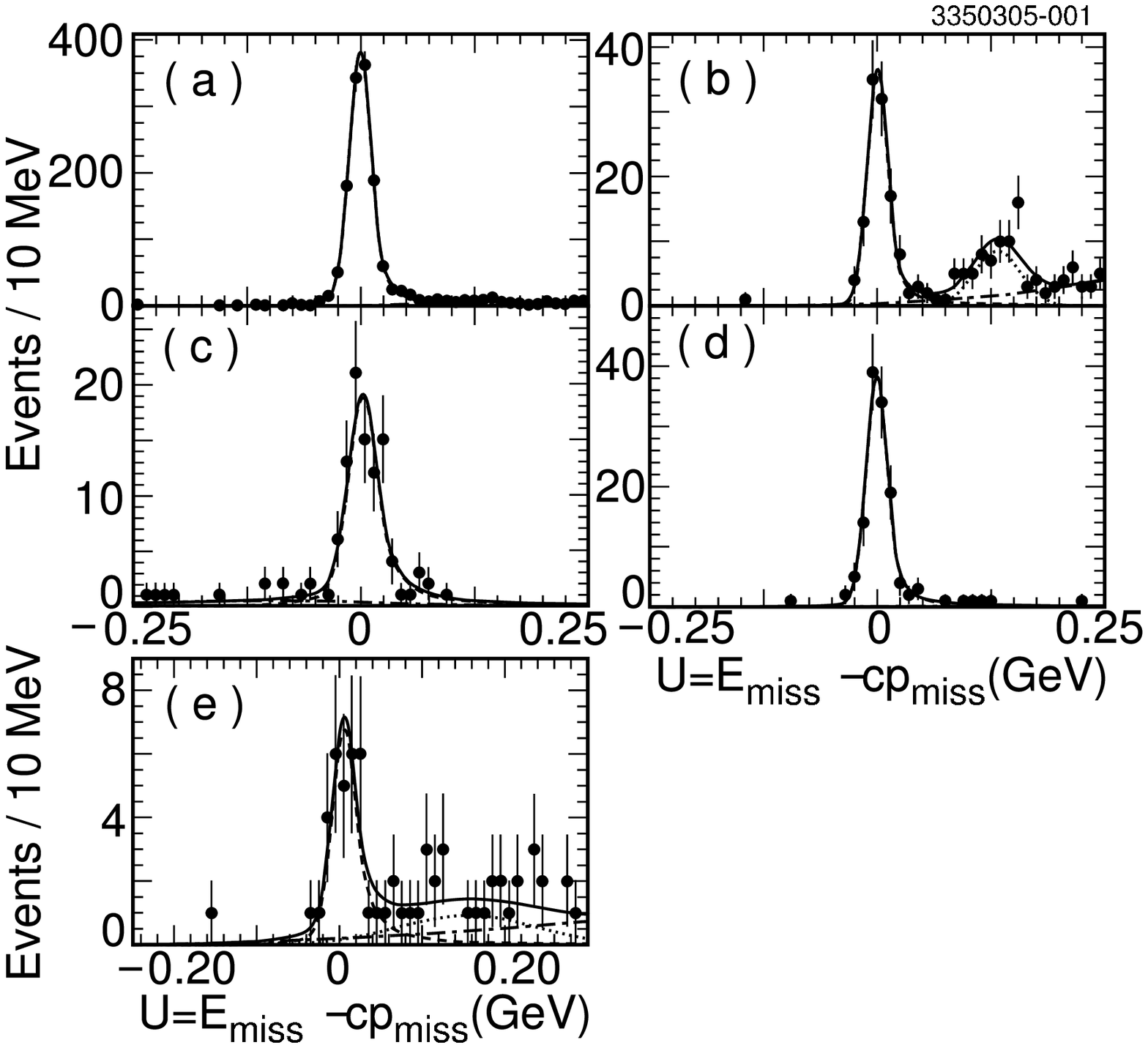}}}
\caption{Fits to $U=E_{\rm miss} - c|\vec{p}_{\rm miss}|$ distributions for 
         (a) $D^0 \to \ke$,  (b) $\pie$, (c) $\kstce$, (d) $\kstne$, and (e) 
         $\rhoe$, with the $\overline {D}^{0}$ meson fully reconstructed.
         The solid line represents the total fit to the data, which includes 
         the signal (dashed line), peaking background (dotted line), and 
         non-peaking background (dot-dashed line).}
\label{signal-yield}
\end{figure}

A tag and the semileptonic decay are then combined.
If there are no tracks other than the daughters of the tag and the 
semileptonic candidate in events, we compute 
$U$, which should peak at zero for a correctly reconstructed
 semileptonic decay. 
To improve the $U$ resolution, we constrain the magnitude of 
the momentum of the $D^0$ candidate decaying semileptonically
 to be $\sqrt{E_{\rm beam}^2/c^2-c^2m_{D^0}^2}$, 
with its direction opposite to that of the tag $\overline D^0$ 
($\hat {\bf p}_{\overline D^0}$)
in the $\psi(3770)$ rest frame, where 
 $\vec{\bf p}_{\overline D^0}=-\vec{\bf p}_{D^0}$. 
The distribution of $U$ is approximately Gaussian
with $\sigma\sim 10$ MeV, varying mode by mode and somewhat larger
for modes with neutral pions.
In case of multiple $K^{*-}$ or $\rho^{-}$ candidates, 
we select a single
combination based on resonance and $\pi^0$ masses.
The $U$ distributions for  
$\dke$, $\pi^- e^+ \nu_e$, $K^{*-} e^+ \nu_e$, and $\rho^{-} e^+ \nu_e$,
summed over tag modes, are shown in Fig.~\ref{signal-yield}.

The signal yields are determined by fitting the $U$ distributions.
The signal is represented with a Gaussian and a Crystal 
Ball~\cite{cb} function to accommodate the tails due to radiative 
effects. The tails of the signal function are fixed to 
the prediction of a {\sc geant}-based~\cite{geant} Monte Carlo (MC) 
simulation.
The background function is determined from MC for each mode.
The backgrounds are small and arise mostly from misreconstructed
semileptonic decays with correctly reconstructed tags. The 
background shape parameters are fixed, while the background 
normalizations are allowed to float in all fits to the data.

The background level for the decay $\dke$ ($\dkste$) 
is small. 
For $\dke$, there is background from $\dkstce$ with an 
undetected $\pi^0$, but it is well separated from the signal because of  
 the missing $\pi^0$.  For the decay 
$\dpie$ ($\drhoe$), the 
background peaks at positive $U$ (well above the signal peak) 
and is primarily from $\dke$
($\dkstce$) where a kaon is misidentified as a pion. 
For $\dkstce$, there is considerable background 
from $\dke$ when a $K^-$ is combined with a random $\pi^0$ candidate, 
which is well below the
signal peak. The signal yield 
for $D^0 \rightarrow K^- e^+ \nu_e$ 
 is determined by separately fitting subsamples for each 
tag mode. The results for all tag modes are 
found to be consistent.
For the other modes, the yields are obtained with all tag modes 
combined due to limited statistics.
The fits to the data are shown in Fig.~\ref{signal-yield}, and the 
yields are given in Table~\ref{table2}. 
The 31.1$\pm$6.3 $\drhoe$ events provide the first observation
of this decay. 

The absolute branching fraction for any $D^{0}$ semileptonic 
decay mode is given by
${\cal B}= N_{\rm signal}/ 
          \epsilon N_{\rm tag}$, 
where $N_{\rm signal}$ is the number of $D^{0} \overline{D}^{0}$ events
with the tag $\overline {D}^{0}$ fully reconstructed and the $D^{0}$
reconstructed in that semileptonic mode, 
$N_{\rm tag}$ is the number of 
$\overline {D}^{0}$ tags, and $\epsilon$ is the effective 
efficiency for detecting the semileptonic decay in an event with an 
identified tag. Note that 
$\epsilon = \epsilon_{\rm signal}$/$\epsilon_{\rm tag}$ is the 
ratio of the separate efficiencies for tag events with semileptonic decays
and tag events in general.  
It is determined with a MC sample that includes relative populations of the
eight tag modes that are consistent with the data.
The cancellation of systematic uncertainties
due to common effects in the numerator and denominator is clear.

We consider the following sources of systematic uncertainty and give
our estimates of their magnitudes in parentheses. 
The uncertainties in the efficiencies 
for finding tracks (0.7\%) and for reconstructing $\pi^0$ (2.0\%) and 
$K^0_S$ (3.0\%) are estimated with missing-mass techniques~\cite{cleoc-Dtagging}
applied to CLEO-c data and MC.  
The uncertainty in the positron-identification efficiency (1.0\%) is 
taken from detailed comparisons of the detector response to positrons of 
radiative Bhabhas in data and MC.  The effect of event complexity was 
incorporated by studying positrons both in isolation and embedded in 
hadronic events.
The positron-identification 
efficiency depends on final-state radiation (FSR) and on 
bremsstrahlung in the material of the CLEO-c detector, the effects of 
both of which are simulated with MC.   To assess the systematic 
uncertainty from these sources (0.6\% combined), we vary the amount of 
FSR (simulated by PHOTOS~\cite{PHOTOS}) and radiation in 
detector material, and we carry out the analyses with and 
without the recovery of radiated photons near positrons.
Uncertainties in the charged pion and kaon identification 
efficiencies~(0.3\% per pion and 1.3\% per kaon) are 
estimated by detailed comparisons of the detector response to tracks from
hadronic $D$-meson decays in data and MC.  There is 
an uncertainty in the number of $\overline {D}^{0}$ tags~(0.7\%), which is 
estimated by using alternative signal functions in the $M_{\rm bc}$ fits 
and by varying the end point (beam energy) of the ARGUS 
function parameterizing the background.
The uncertainty in modeling the background shapes in the $U$ fits 
(mode dependent: from 1.0\% to 5.0\%) has contributions from 
the simulation of the positron and hadron misidentification rates, 
as well as the input branching fractions. 
The uncertainty associated with the requirement of no extra tracks in 
a candidate event (0.5\%) is estimated using fully reconstructed 
$D^{0} \overline{D}^{0}$ events in the data and MC. The uncertainty in the 
semileptonic reconstruction efficiencies due to imperfect
knowledge of the semileptonic form factors is small because of the 
uniform acceptance of the CLEO-c detector. It is estimated by varying the 
form factors in the MC within their uncertainties 
(1.0\%) for all modes except $D^0 \rightarrow \rho^- e^+ \nu_e$, 
for which a conservative uncertainty (3.0\%) is used in the absence of 
experimental information on the form factors 
in Cabibbo-suppressed pseudoscalar-to-vector transitions. 
The uncertainty associated with the simulation
of initial state radiation~($e^+ e^- \rightarrow D \overline{D} \gamma$) is
found to be negligible. Finally, there is systematic uncertainty 
due to limited MC statistics (0.7\% to 1.5\%, depending on mode).

A non-resonant component is likely to contribute background in 
semileptonic decays to vector mesons. 
Based on evidence from the FOCUS experiment of
 a non-resonant component consistent with an S-wave 
interfering with $D\to K^{*}\ell\nu_l$~\cite{focus}, its 
contribution is estimated to be 2.4\% in this analysis and  
subtracted when calculating the branching fractions for 
$\dkste$. Systematic uncertainty associated with 
the subtraction (1.0\%) is due to imperfect knowledge of 
the amplitude and phase of the non-resonant component. 
Interference with the S-wave amplitude 
alters the angular correlations among the decay products and 
introduces a systematic uncertainty  (1.5\%) in the reconstruction 
efficiency for $\dkstce$. 
A relativisitic Breit-Wigner with a Blatt-Weisskopf
form factor is used to simulate wide resonances in MC.   
A systematic uncertainty associated with the $K^{*}$ lineshape  (1.2\%) 
is estimated using 
$D^+\rightarrow {\overline K^{*0}}(K^-\pi^+)e^+\nu_e$, which has 
a much larger yield~\cite{Dplus}. 
 For $D^0\rightarrow \rho^-e^+\nu_e$, there are insufficient data to
constrain the non-resonant background or the resonance lineshape. 
Systematic uncertainties from these two sources 
 are expected to be much smaller than the
current statistical uncertainty for this mode, and they are neglected.

\begin{table*}[htbp] 
\caption{ Signal efficiencies, yields with statistical uncertainties, 
         and branching fractions with both statistical and systematic 
         uncertainties  
         for $\dke$, $\pi^- e^+ \nu_e$, $K^{*-} e^+ \nu_e$, and 
         $\rho^{-} e^+ \nu_e$, as well as 
         PDG branching fractions with total 
         uncertainties~\cite{PDG} and 
         BES branching fractions  with statistical 
         and systematic uncertainties~\cite{PLB597}. 
         For $\dkste$, the efficiencies do not include 
         subsidiary decay branching fractions. 
         The branching fractions for $D^0 \rightarrow  {K}^{*-} e^+ \nu_e$ 
         are reduced by 2.4\%~(see the text). }
\begin{center}
\begin{tabular}{lcccccc}\hline  \hline
Decay Mode  & $\epsilon$ (\%)   & Yield    & ${\cal B}$(\%)  & ${\cal B}$(\%) (PDG) 
            & ${\cal B}$(\%) (BES)                                    \\ \hline 
$\dke$    & 63.58$\pm$0.50              & 1311.0$\pm$36.6  &
            3.44$\pm$0.10$\pm$0.10     &   3.58$\pm$0.18  & 
            3.82$\pm$0.40$\pm$0.27                                     \\ 
$\dpie$   & 74.18$\pm$0.52              &  116.8$\pm$11.2  &  
           0.262$\pm$0.025$\pm$0.008   &   0.36$\pm$0.06  &  
            0.33$\pm$0.13$\pm$0.03                                     \\ 
$\dkstce$ & 22.02$\pm$0.32              &   94.1$\pm$10.4  &
            2.11$\pm$0.23$\pm$0.10     &                  &            \\ 
$\dkstne$ & 40.43$\pm$0.42              &  125.2$\pm$11.6  & 
            2.19$\pm$0.20$\pm$0.11     &                  &            \\ 
$\dkste$  & & & 2.16$\pm$0.15$\pm$0.08  &  2.15$\pm$0.35   &            \\ 
$\drhoe$  &  26.97$\pm$0.35 & 31.1$\pm$6.3 & 0.194$\pm$0.039$\pm$0.013 & &
 \\  \hline \hline
\end{tabular}
\end{center}
\label{table2}

\caption{Ratios of branching fractions for 
         exclusive $D^0$ semileptonic decays, from this analysis, the 
         PDG~\cite{PDG}, CLEO III~\cite{hsu}, and FOCUS~\cite{RFocus}.
         Uncertainties in the CLEO and FOCUS measurements are statistical and 
         systematic, respectively.}
\begin{center}
\begin{tabular}{lccc}\hline \hline
                     &  {${\cal B}(\dpie)\over{\cal B}(\dke)$}
                     &  {${\cal B}(\drhoe)\over{\cal B}(\dkste)$}     
                     &  {${\cal B}(\dpimu)\over{\cal B}(\dkmu)$}       \\ \hline
This measurement     & (7.6$\pm$0.8$\pm$0.2)\% 
                     & (9.0$\pm$1.9$\pm$0.6)\%   &                     \\ 
PDG~\cite{PDG}       & (10.1$\pm$1.8)\%
                     &                           &                     \\ 
CLEO III~\cite{hsu}  & (8.2$\pm$0.6$\pm$0.5)\%  
                     &                           &                     \\ 
FOCUS~\cite{RFocus}  &                           &
                     &  (7.4$\pm$0.8$\pm$0.7)\%                        \\ \hline \hline 
\end{tabular}
\end{center}
\label{summary}

\end{table*}

The estimated systematic uncertainties are added in quadrature to 
obtain the total systematic uncertainties in the branching fractions
(Table~\ref{table2}): 2.8\%, 2.9\%, 4.9\%, 
5.1\%, and 6.6\% for $\dke$, 
$\dpie$, $\dkstce$, $\dkstne$, and $\drhoe$, respectively.  Most of
the estimates of systematic uncertainties are limited by data statistics 
and will be reduced with a larger data sample.

In summary, we have presented absolute branching fraction 
measurements of $D^{0}$ semileptonic 
decays 
with the first 55.8~pb$^{-1}$ of data collected with
the CLEO-c detector at the $\psi(3770)$.  Our 
branching fractions for $D^0$ decays to $\ke$, $\pie$, 
$\kste$, and $\rho^{-} e^+ \nu_e$
are given in Tables~\ref{table2} and ~\ref{summary}, along with 
current world-average values compiled by the Particle Data 
Group~\cite{PDG}. Recent CLEO III~\cite{hsu}, BES~\cite{PLB597} and
FOCUS~\cite{RFocus} measurements, which are
not included in Ref.~\cite{PDG}, are also listed. 
The measurement of $\drhoe$ is the first observation of this mode.  
The absolute branching fraction measurements of 
other modes are more precise than 
and consistent with current world averages.
Corresponding results for $D^+$ semileptonic decays and more extensive
interpretation are presented in a companion  Letter~\cite{Dplus}.


We gratefully acknowledge the effort of the CESR staff 
in providing us with excellent luminosity and running conditions.
This work was supported by the National Science Foundation
and the U.S. Department of Energy.

\end{document}